# Trends in Combating Image Spam E-mails

Mohammadi Akheela Khanum [a]**,** Lamia Mohammed Ketari [b]

Information Technology Department, College of Computer and Information Sciences

King Saud University, Kingdom of Saudi Arabia

[a]kakheela@ksu.edu.sa, [b]lketari@ksu.edu.sa

**Abstract.** With the rapid adoption of Internet as an easy way to communicate, the amount of unsolicited e-mails, known as spam e-mails, has been growing rapidly. The major problem of spam e-mails is the loss of productivity and a drain on IT resources. Today, we receive spam more rapidly than the legitimate e-mails. Initially, spam e-mails contained only textual messages which were easily detected by the text-based spam filters. To evade such detection, spammers came up with a new sophisticated technique called image spam. Image spam consists in embedding the advertisement text in images rather than in the body of the e-mail, yet the image contents are not detected by most spam filters. In this paper, we examine the motivations and the challenges in image spam filtering research, and we review the recent trends in combating image spam e-mails. The review indicates that spamming is a business model and spammers are becoming more sophisticated in their approach to adapt to all challenges, and hence, defeating the conventional spam filtering technologies. Therefore, image spam detection techniques should be scalable and adaptable to meet the future tactics of the spammers.

**Introduction**

Malicious software, also called malware is a software that enters a computer system without the user's knowledge or consent. Mark Ciampa [1] has divided the malware based upon their primary objectives, into (i) infecting malware (viruses and worms), (ii) concealing malware (trojan horses, rootkits and logic bombs), (iii) malware for profits (spam, spyware and botnets). Spam also known as junk, is defined as unsolicited e-mails sent by spammers to increase sales of their cut-rate products (e.g: counterfeit products, pharmaceuticals, fake diplomas, etc). The volume of spam has extremely grown in the last few years because of the increased use of Internet. In fact, marketers are bringing e-commerce to inboxes mainly by using Internet as an effective and less expensive medium. Usually, spammers collect e-mail addresses from chat rooms, websites, customer lists, newsgroups, and viruses which harvest users' address books. Sending spam is a lucrative business since spammers don't have to bear any cost in sending millions of spam e-mail messages. If they receive responses from a very small percentage of receivers, they can make millions of dollars. According to *Postini* [2], a communications and security compliance firm, one out of every 12 e-mails is spam. Spam significantly reduces work productivity. Infact, more than 11% of workers receive 50 spam messages each day and spend more than half an hour deleting them [2]. Spam e-mails can be classified into six broad categories as follows:

- Advertisement
- Adult content
- Lottery winning notification
- Health and Pharmacy
- Online Degrees
- Bank and Finance

To block spam, often service providers and companies relied upon tracing some keywords which were very frequently used in the spam, such as the word 'viagra' or 'bank'. However, adding more keywords could result in false positives and would block legitimate e-mail. Another way of blocking spam included, making use of blacklists that contain a list of IP addresses of known spammers or compromised hosts. However, these lists have to be constantly updated because the spammers have learnt to counteract this by rapidly changing the origin of spam [3]. Since the text-based spam filters are increasingly being used to filter text-based spam, the spammers have come up with a new approach to send their spam: the image spam. Spammers inlcude their advertisements as part of an embedded image file attachment (.gif, .jpg, .png, etc) rather than the body of the e-mails, hence defeating the text-based spam filtering techniques. Image spam may be a single image or it can contain several images within a single image. Image spam may also use word splitting which involves horizontally separating words, although still readable by the human. Some examples of the image spam are shown in Fig. 1.

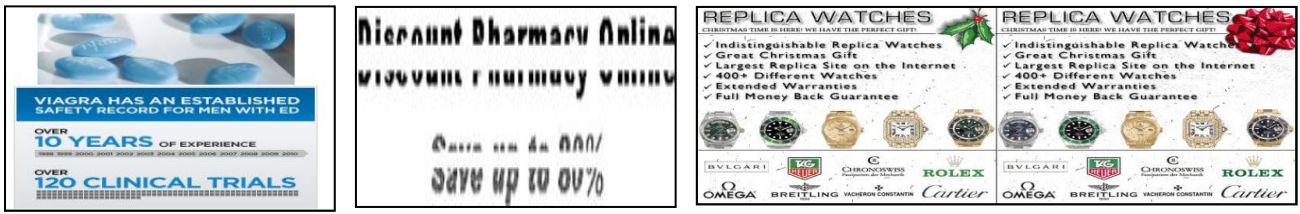

Fig. 1: Examples of image spam

Image spam exploded in 2006 and by early 2007 it had reached a peak of over 50% of total spam received [4] and its menace is still going. Fig. 2 gives the image spam spread rate from May 2010 till March 2011 according to *M86 Security Labs*[5].

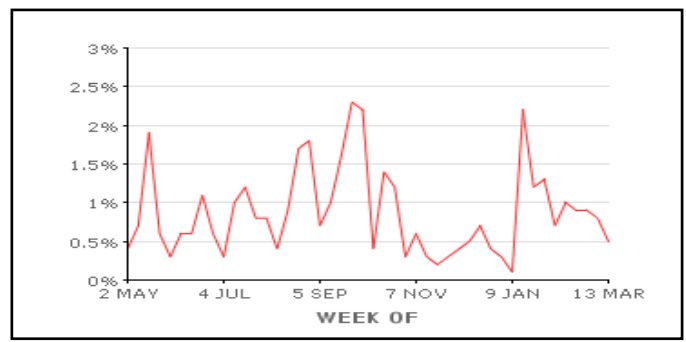

Fig. 2: Image Spam spread rate

**Motivations and Background**

Global spam volume increased very fast over the past five years. E-mail spam accounted for 96.5% of incoming e-mails received in business by June 2008 [6]. *Nucleus research* reports that spam e-mail, on average, costs U.S organizations $874 per person annually in lost productivity [2]. The success of text-based spam filtering techniques has driven spammers to find new variations of spam, and their latest invention is the image spam. As reported by *McAfee* [7], image spam accounts for approximately 30% of all e-mail spam. The main objectives of the spammers are:
- *To generate profit in the form of money:* spammers send advertisements embedded in the spam e-mails.
- *To promote products and services:* companies deal with the spammers and pay them to promote their products and services.
- *To steal sensitive information such as credit card numbers, passwords and bank account details*: this may be achieved in two ways (i) back door entry created by malicious programs and (ii) launching phishing attacks.

E-mail spam is a growing and serious problem faced by e-mail administrators and users. In early years of the past decade, e-mail spam was adequately handled by text filtering techniques, since spam were text-based. Text-based detection techniques, such as heuristics and Bayesian filters, were widely used and revealed to be efficient core techniques to handle text-based spam. More recently the high increase in the quantity of spam can be attributed to more advanced techniques such as the image spam. Early image spam simply embedded advertising text in images that linked to HTML formatted e-mail. The contents could be automatically displayed to end-users while being shielded from text-based spam filters [8]. To deal with image spam, filtering technologies began to incorporate Optical Character Recognition (OCR)[1] into the filters to detect the text in the images. Spammers then moved one step ahead and began applying CAPTCHA (Computer Automated Public Turing Test to Tell Computers and Humans Apart) techniques. These techniques distort the original image or add colors or noisy background so that only humans can identify the intended message [9]. Since then, many image spam detection techniques where designed to effeciently separate spam images from non spam images. Image classification techniques where employed to do this. Classification is done based on several image properties also called as image features. Image features can be either high-level or low-level. High-level image features comprise of the information in the image header such as the file name, file format, aspect ratio, image area, compression, horizontal, vertical resolution etc. Low-level image features (visual features) comprise of color features, texture features, shape features and appearance features.

In what follows, we enumerate the main image spam detection techniques found in the literature.

**Related Work**

Dredze et al. [10] propose fast classifiers by introducing Just in Time (JIT) feature extraction, which creates features at classification time as needed by the classifier. The proposed technique focuses on the simple properties of the image rather than using a complex analysis. In addition to using basic features, the technique also uses some advanced features such as image size, color, edge detection and random pixel test. The technique is tested using maximum entropy which has comparable state of the art performance with Support Vector Machine (SVM) on a wide range of tasks. Evaluations on data reflecting a real world distribution over spam images yielded up to 97% accuracy.

Krasser et al. [11] came up with a framework for feature extraction and classification that operates on features that can be extracted from image files in a very fast fashion. They used four basic image features namely the width, the height, the image file type, and the file size, which can be quickly derived from an image at an extremely low computational cost. They used two popular classification algorithms, the C4.5 algorithm for building decision tree and the SVM algorithm for building SVM. The evaluation results shows that about 60% of the image spams can be eliminated using the proposed technique with low false positive rate of 0.5%. Therefore, the proposed low-cost classification can be used as the first tier in a multi-tier classification framework to eliminate large amount of spam images before doing expensive calculations.

Bhasker et al.[12] present two techniques for image spam detection. The first uses the visual features such as the color, texture and shape of the image for classification using SVM. This classification offers 95% accuracy in all cases. The second technique uses near duplicate detection in images. It involves clustering of image Gaussian Mixture Models (GMM) based on the Agglomerative Information Bottleneck (AIB) principle. GMM based labeled AIB has high accuracy of 93% when predicting spam in comparable conditions as compared to OCR and trained SVM. Overall, the proposed technique gives a prediction accuracy of 95%.

Wan et al. [13] propose an effective algorithm called Edge Classification-Based Text Localization (ECTL) to extract the text regions in spam images. The algorithm consists of four stages: (i) edge detection, (ii) corner detection, (iii) edge classification, and (iv) candidate text regions refine. It also uses color edge detector to extract edges of the color images. The experimental results shows that the algorithm can identify 96% of texts contained in spam images and the precision can reach up to 97.6% on real world data.

---
[1] OCR is the mechanical or electonic translation of scanned images of handwritten, typewritten or printed text into machine-encoded text.

Uemura and Tabata [14] propose a technique which allows the existing Bayesian filter to learn image information such as file name, file size, area of image, and compressibility. The technique is applied to GIF image types, which accounts for majority of image spam. The proposed technique was implemented on Bayesian filter which learns image information. The experimental results indicates that the proposed technique can realize a false negative rate lower than that of the conventional Bayesian filter technique.

Zang et al. [15] suggest a multimodal framework to reveal the source of image spam through three steps: (i) image segmentation, (ii) feature extraction and similarity calculation, and (iii) image spam clustering. The framework uses color features, layout features, texture features and text content for image classification. Based on the extracted features, spam images are categorized into four different types: the first one containing mainly text, the second type containing mainly foreground illustrations, the third type containing a mixture of text and illustrations, while the fourth type neither text nor illustrations. Conditional entropy-based technique is used to evaluate the clustering results, which shows that the technique gives better results as compared to other similar techniques.

Cheng et al. [16] propose a framework called Binary Filtering with Multi-Label Classification (BFMLC) to take both spam image filtering and user preferences into account. BFMLC comprises of two-stage classification: filter-oriented binary classification and user-oriented multi-label classification. A file based on the BFMLC framework cannot only discriminate spam image from non spam images (also called ham images) but also classifies spam image as several predefined topics. BMLC framework was tested on public personal datasets. The experimental results shows that the system can identify spam images with the average accuracy of 96.30% and classify spam images as predefined topics with the average precision of 89.42%.

The technique proposed by Liu et al. [17] comprise of a three-layer image spam filtering system is proposed. The system filters the image spam by analyzing both the mail header and the image. The first layer of the system extracts the header features, and then the mails are filtered by Bayesian Classifier. The second layer focus on analyzing the image content, first the high-level features such as the file name, width, file size, aspect ratio, height, and image format, are extracted. These features are passed through the SVM classifier to decide whether the image is a spam or not. If the first two layers do not have a consistent result, the system makes the final decision by analyzing the low-level features of the image such as the color histogram and the color moment. The experimental results shows that most image spam mails are identified by the first layer and the accuracy rate of 94% is achieved for the whole system.

Wang et al. [18] propose a feature extraction scheme that focuses on low-level features which includes the metadata and visual features. Metadata features included the image size, width, height, bit dept, and image file type. The visual features included the color features such as the number of colors, variance, mostly appearing color, primary color and color saturation, the texture feature which is represented by the histogram. Based on these extracted properties of image, a one-class SVM classifier with radial basis function (RBF) kernel as the kernel function is used to classify image as ham or spam. The experimental results shows that this technique provides a detection rate of 95% for various datasets.

Gao et al. [19] propose a comprehensive solution for image spam detection which performs both server-side filtering and client-side detection to effectively mitigate image spam. On the server-side, a non-negative sparsity induced similarity measure for cluster analysis of spam images to filter the attack activities of spammers and fast trace back the spam source is used. At the client-side, the technique uses the principle of active learning where the learners guide the users to label as few images as possible while maximizing the classification accuracy. The results indicated that the standard deviations of the performance quantities of the proposed approach are smaller than those of the competition techniques, which is an indication that the proposed measure is more preferable. At the client-side, the SVM achieved 99% recognition accuracy.

More recently, the work proposed by Soranamageswari and Meena [20] involves a technique in which the gradient histogram is used as the key feature. The gradient values are evaluated for each pixel of an image. The obtained features are then normalized for efficient spam classification. The normalized features are then applied as input for feed forward back propagation neural network (BPNN) model which classifies the image spam from those of legitimate mail. The average classification accuracy of around 93.7% is obtained on 9/10 training and testing sets.

**Discussion**

Since 2006, many researches focused on finding novel and efficient techniques to combat the image spam. Some techniques try to analyze the image content to extract text, while others focus on visual properties such as color, texture and variance of image. The image spam filtering is viewed as a kind of classification problem. Well-known classifier algorithms such as the Naïve Bayesian Classifier [14,17], Support Vector Machine (SVM) [11,12,17,18] as well as Neural Networks [20], are extensively being used to design efficient image spam filtering techniques. Various image and textual features are employed for the classification of image spam. Image features usually used are the high-level image features and low-level image features. High-level image features are easier to analyze and has low processing cost as compared to low-level image features. Some techniques also use text features such as the edge pattern and edge density for image spam detection.

The study and the anlysis of various techniques for image spam detection in the previous section, guided us to produce Table 1 which summarizes image spam detection approaches based upon the classification techniques and the type of features to distinguish a spam and a ham.

Table 1: Overview of image spam detection techniques

| Related Work | Technique used for image spam filtering/classification | Features used for classification | | | Filtering/Classification Results |
| --- | --- | --- | --- | --- | --- |
| | | Image features | | Text features | |
| | | low-level | high-level | | |
| [10] | Just in Time (JIT) feature extraction | √ | √ | X | **97%** |
| [11] | C4.5 Decision trees and Support Vector (SVM) Machine learning | X | √ | X | **60%** |
| [12] | SVM and Gaussian Mixture Models (GMMs) | √ | X | X | **95%** |
| [13] | Edge Classification Based Text Localization | X | √ | √ | **97.60%** |
| [14] | Bayesian Filtering | X | √ | X | **90.00%** |
| [15] | Multimodal Clustering | √ | X | √ | **97.20%** |
| [16] | Multi-label k-nearest neighbor(ML-KNN) | √ | X | X | **96.31%** |
| [17] | Baysian classifier and SVM classifier | √ | √ | X | **94%** |
| [18] | SVM with Radial Basis Function(RBF) kernel | √ | X | X | **95%** |
| [19] | Nonnegative sparsity induced similarity measure and principle of active learning | √ | X | X | **99%** |
| [20] | Feed forward Back Propagation Neural Network (BPNN) | √ | X | X | **93.70%** |

Most image spam detection techniques include automatic pattern classifiers based on machine learning or some type of pattern matching. These machine learning techniques can be either supervised or semi-supervised. Supervised learning algorithms require large amount of training data set of labeled e-mails to obtain better boundaries between an image spam and non image spam e-mails. Supervised learning also requires a frequent update because of the changing ways of the spammers to evade filters. The techniques listed in Table 1 uses supervised learning for training the classifiers. Few studies [21] have used semi-supervised learning techniques for image spam detection. Semi-supervised learning addresses a common issue that arises with many supervised learning works, the issue of collecting a large set of labeled data, but unlabeled data sets is easier to obtain. Semi-supervised techniques can use small set of labeled examples along with a larger set of unlabeled examples in order to train the classifiers. However, the obtained classfication results depend on both the choice of the features and also the technique used for classification. Low-level image features gives better insight to the classification as compared to using only high-level image features. Therefore, using appropriate combination of the features and the classification algorithm is preferred. We also found that some filtering/classification techniques gives high accuracy of detection but under predifined conditions, either the results are obtained only for a predefined corpus or in some cases the accuracy is obtained for only one type of image spam like the .gif images [14]. In addition, some of these techniques are applicable to detect only predefined obfuscations for the images.

**Conclusion**

The industry is seeing a rise in spam attacks where many existing e-mail filtering technologies are no longer effective. Effective technologies and organizations that have the skills to understand and deal with these attacks are needed. The recent emergence of image spam has presented new challenges,some of the most prominent of them are:
- Image spam detection is more difficult and it may require more time to upgrade the conventional e-mail filters in order to detect image spam.
- Image spam message is usually significantly larger than a text-based spam message. In fact, a text-based spam message is approximately 5 KB to 10 KB in size while the size of the image spam message is approximately 10 KB to 70 KB. Therefore, it will be required to allocate significantly more bandwidth and storage for e-mail infrastructure. The solution is not as easy because stopping to examine the image files means a jam at the mail server which can also lead to lost e-mail. Therefore, many organizations feel it is better to stop the spam at the firewall before the mail server and this can also save the bandwidth.
- Image spam messages take longer to analyze because they are considerably larger in size than text-based spam, and the analysis is time consuming and may create chaos at the mail server. It may be required to upgrade or replace existing platforms to cope with the needed processing power

New obfuscation techniques are implemented in real-time and early generation technologies and techniques are not able to cope with the shifting nature of attacks. Spammers no more rely on one technique, and consequently, image spam has become a latest trend in the fight between spammers and e-mail users. The techniques that we surveyed are efficient in detecting image spam but under limitations. In fact, some work well with a particular type of images, others give good detection when combined with multiple layers of filtering. Also, some techniques takes lot of processing time. Because spammers constantly keep changing their tactics, spam filtering solutions must have flexible design to be able to meet today's threat and adaptable to future strategies and image spam detection techniques should be scalable to meet the demands of the real world problems (mainly, loss of time and loss of productivity). To conclude, this study has formed the basis for our future work in image spam detection . Also, it has established the line of investigation that is needed to move forward in designing and implementing a new framework dedicated to efficient protection against image spam.